\documentclass[final,3p]{elsarticle}

\usepackage[]{graphicx}
\usepackage{amssymb}
\usepackage{amsmath}
\usepackage{array}
\usepackage{multicol}
\usepackage{color}
\usepackage{lineno}

 \biboptions{square,comma}

\journal{J.~Comput.~Phys.}

\begin{document}

\begin{frontmatter}

\title{Stochastically generated turbulence with improved kinematic properties}


\author[DTU]{Mads M{\o}lholm Hejlesen}
\author[DTU,ETH]{Jens Honor\'{e} Walther\corref{cor1}}

\address[DTU]{Department of Mechanical Engineering, 
Technical University of Denmark, Building 403, DK-2800 Kgs.\ Lyngby, Denmark}
\address[ETH]{Computational Science and Engineering Laboratory, 
ETH Z\"{u}rich, Clasiusstrasse 33, CH-8092 Z\"{u}rich, Switzerland}

\cortext[cor1]{Corresponding author at: 
Department of Mechanical Engineering, Technical University of Denmark,
Building 403, DK-2800 Kgs.\ Lyngby, Denmark. 
Tel.: + 45 4525 4327; fax: + 45 4588 4325. 
E-mail address: jhw@mek.dtu.dk (J.~H.~Walther).}

\begin{abstract}
We present a stochastic turbulence generator based 
on a vorticity formulation where the generated turbulent field implicitly 
fulfills the kinematic constraints of an incompressible flow. 
The generator allows direct access to the turbulent 
velocity and vorticity field. 
Enforcing additional constraints such as a divergence-free vorticity field 
and a specified differentiability of the flow field 
can also implemented directly within this formulation.
The resulting turbulent field contain improved kinematic properties and 
may be imported into numerical simulations 
without an excessive loss of energy.
\end{abstract}

\begin{keyword}
Turbulence generator \sep 
Stochastic simulations \sep 
Synthesized turbulence \sep
Vorticity formulation
\end{keyword}

\end{frontmatter}

\section{Introduction}

A stochastic turbulent field is a randomly generated field that 
possess the same statistical correlation properties as a corresponding 
real turbulent field but is generally not a solution to the governing 
equations of fluid flow i.e.\ the Navier-Stokes equation.
Stochastic turbulence generators are used in fluid dynamics simulations 
to emulate the effect of turbulence in the oncoming flow. A common use is 
simulating a fluid-structure interaction with a turbulent oncoming flow 
in order to investigate the effect of high-frequency fluctuations 
on the aero-elastic coupling 
\citep{Prendergast:2006,Prendergast:2007,Rasmussen:2010,Hejlesen:2015b}. 
Stochastic turbulence generators are also being used in large eddy simulations 
for generating non-deterministic boundary conditions for the resolved 
(i.e. simulated) turbulent scales \citep{Sagaut:1998}.

Current stochastic turbulence generators \citep{Mann:1998,Prendergast:2007} 
are based on the method of Shinozuka \citep{Shinozuka:1972} for simulating 
a statistically homogeneous process of the velocity field. 
However creating the velocity field by a stochastic process does not 
generally comply with the kinematic constraints of a flow field.
Hence the generated turbulent energy is distributed into the components of an 
unconstrained velocity field. 
Consequently a significant part of the kinetic energy may be lost in a 
numerical simulation when applying kinematic constraints on the unconstrained 
velocity field 
e.g.\ enforcing the divergence-free velocity field of an incompressible flow.

We show that by formulating the stochastic method in a vorticity formulation 
some of these constraints are implicitly fulfilled and additional 
constraints may be enforced by a re-projection method.
This approach ensures that the kinetic energy is contained 
in components which comply to the kinematics of the flow. 
Furthermore, we discuss how to improve the numerical and physical properties 
of the generated field to ensure that it can be introduced into a 
numerical simulation without having an excessive loss of turbulent energy. 



\section{Methodology}


A vector field $\boldsymbol{v}$ that has at least 
two continuous derivatives (i.e.\ $\boldsymbol{v} \in C^2$) can be viewed as 
consisting of two intrinsic components according to the Helmholtz decomposition:  
\begin{equation}
 \boldsymbol{v} = \boldsymbol{\nabla} \times \boldsymbol{\psi} - \boldsymbol{\nabla} \phi
\qquad \mbox{with} \qquad
 \boldsymbol{\nabla} \cdot \boldsymbol{\psi} = 0
 \label{eq-4:Helmholtz}
\end{equation}
Here $\phi$ is a scalar potential and $\boldsymbol{\psi}$ is a vector potential which 
is divergence-free in order to ensure the uniqueness of the decomposition. 
The two terms represents a divergence-free and a irrotational component of 
the velocity field, respectively. 
For a constant density fluid the conservation of mass imposes the 
constraint of a divergence-free velocity field 
$\boldsymbol{\nabla} \cdot \boldsymbol{v} = 0$. 
Hence the irrotational component of the velocity 
field is nullified $\boldsymbol{\nabla} \phi = \boldsymbol{0}$ and 
the velocity field can be formulated by the vorticity field alone.
Using the definition of vorticity 
$\boldsymbol{\omega} \equiv \boldsymbol{\nabla} \times \boldsymbol{v}$ we 
thus obtain the relation:
\begin{equation}
 \boldsymbol{\omega} = \boldsymbol{\nabla} (\boldsymbol{\nabla} \cdot \boldsymbol{\psi}) - \boldsymbol{\nabla}^2 \boldsymbol{\psi}
         = - \boldsymbol{\nabla}^2 \boldsymbol{\psi}
\qquad \mbox{by which} \qquad
 \boldsymbol{\nabla} \times \boldsymbol{\omega} = - \boldsymbol{\nabla}^2 \boldsymbol{v}
 \label{eq-4:kinematics}
\end{equation}
Hence the velocity field can be obtained from the vorticity field by solving a 
Poisson equation. We see that by generating a turbulent vorticity field, 
from which the velocity field is then calculated, the 
constraint imposed by mass conservation is implicitly fulfilled, 
and the turbulent energy 
is thus fully contained in a divergence-free velocity field. 
This is not the case when generating the velocity field directly and 
supplementary numerical schemes must then be used to enforce this constraint.

%

A correlated random vector field can be generated by convolving a random white 
noise vector field $\boldsymbol{\phi}$ of unit variance with the desired covariance 
tensor function $\boldsymbol{H}$:
\begin{equation}
 \boldsymbol{\omega}(\boldsymbol{x}) = \int \boldsymbol{H} (\boldsymbol{x} - \boldsymbol{y}) 
  \cdot \boldsymbol{\phi}(\boldsymbol{y}) \ d\boldsymbol{y}
 \qquad \mbox{and thus} \qquad 
 \widehat{\boldsymbol{\omega}}(\boldsymbol{k}) = 
     \widehat{\boldsymbol{H}} (\boldsymbol{k}) \cdot 
     \widehat{\boldsymbol{\phi}}(\boldsymbol{k})
\label{eq-4:generation}
\end{equation}
Here $\widehat{\,\cdot \,}$ denotes the Fourier transform and $\boldsymbol{k}$ 
is the angular the wave-number. In this work the multi-variate formulation 
simply represents the components of the vorticity vector. 
However the multi-variate formulation 
may also be used to introduce a statistically inhomogeneous direction in the 
turbulence field c.f.\ \citep{DiPaola:1998}.
Furthermore, for a turbulent field which is advected with a constant velocity $U$ 
(e.g.\ in the $x$-direction) we may use Taylor's frozen turbulence assumption to 
transform a spatial dimension into a temporal one 
e.g.\ by $\boldsymbol{x} = (U \, t,y,z)$.

In order to find an expression for the covariance tensor $\boldsymbol{H}$ we start 
with the expression for the spectral power density tensor of the turbulent 
vorticity field $\widehat{\boldsymbol{\Omega}}$ combined with the spectral 
correlation properties of the components 
of the random vector field $\boldsymbol{\phi}$:
\begin{equation}
 \widehat{\boldsymbol{\Omega}} = 
\frac{ \widehat{\boldsymbol{\omega}} \ \widehat{\boldsymbol{\omega}}^* }{V}
    = \frac{ (\widehat{\boldsymbol{H}} \cdot \widehat{\boldsymbol{\phi}} ) 
           \ ( \widehat{\boldsymbol{H}} \cdot \widehat{\boldsymbol{\phi}} )^* }{V}
\qquad \mbox{with} \qquad 
  \widehat{\boldsymbol{\phi}} \ \widehat{\boldsymbol{\phi}}^* = \boldsymbol{I}
\end{equation}
As $\boldsymbol{\Omega}(\boldsymbol{k})$ is a symmetric matrix 
consisting of real, positive and even functions 
of $\boldsymbol{k}$ we get:
\begin{equation}
 V \, \widehat{\boldsymbol{\Omega}} = 
  \widehat{\boldsymbol{H}} \widehat{\boldsymbol{H}}^\dagger
\end{equation}
Here the right-hand-side is a matrix multiplication and the $\dagger$ denotes 
the conjugate transpose by which we see that the covariance tensor 
function $\widehat{\boldsymbol{H}}$ may be found by a Cholesky decomposition 
of $\widehat{\boldsymbol{\Omega}}$. 
The turbulent vorticity field can thus be generated by Eq.~\eqref{eq-4:generation} 
given a well-defined $\widehat{\boldsymbol{\Omega}}$.

We may relate the vorticity power density function to the kinetic energy of 
the velocity field by specifying the velocity field as an even function i.e. 
$\widehat{\boldsymbol{v}}^*(\boldsymbol{k}) = \widehat{\boldsymbol{v}}(\boldsymbol{k})$,
by which the vorticity field effectively becomes an odd function i.e.\ 
$\widehat{\boldsymbol{\omega}}^*(\boldsymbol{k}) = 
  -\widehat{\boldsymbol{\omega}}(\boldsymbol{k})$. 
This specification formally changes the periodic boundary condition of the velocity 
field to a symmetrical one.
The spectral vorticity power density tensor function may thus be obtained:
\begin{equation}
 \widehat{\boldsymbol{\Omega}}(\boldsymbol{k}) = - \frac{\widehat{\boldsymbol{\omega}} \ \widehat{\boldsymbol{\omega}}}{V} 
    = \frac{ |\widehat{\boldsymbol{v}}|^2  (|\boldsymbol{k}|^2 \boldsymbol{I} - \boldsymbol{k} \, \boldsymbol{k}) 
    - |\boldsymbol{k}|^2 \ \widehat{\boldsymbol{v}} \, \widehat{\boldsymbol{v}} }{V}
\end{equation}
Here we have used the spectral definition of 
vorticity $\widehat{\boldsymbol{\omega}} = \iota \, \boldsymbol{k} \times \widehat{\boldsymbol{v}}$ 
and the incompressibility constraint 
$\iota \, \boldsymbol{k} \cdot \widehat{\boldsymbol{v}} = 0$.
By further assuming an isotropic turbulence, the spectral vorticity power density 
tensor may be stated as:
\begin{equation}
 \widehat{\boldsymbol{\Omega}}(\boldsymbol{k}) = \frac{E(|\boldsymbol{k}|)}{4 \pi \, \rho \, |\boldsymbol{k}|^2}(|\boldsymbol{k}|^2\boldsymbol{I} - \boldsymbol{k} \, \boldsymbol{k})
 \label{eq-4:isotropic}
\end{equation}
where the $E(|\boldsymbol{k}|)$ is the kinetic energy spectrum 
(Eq.~\eqref{eq-4:isotropic} is also given by \citep{Batchelor:1953}).

%

The resulting field that is generated by Eq.~\eqref{eq-4:generation}, is at 
this point a pure Monte Carlo method and thus powered by a 
random number generator. 
Consequently, the generated field does not posses any properties which 
allows the field to comply to a numerical simulation of differential equations, 
e.g.\ smoothness.
Evidently this non-compliance will result in a large numerical dissipation and 
thus loss of kinetic energy when introducing the turbulent field in a numerical 
simulation.
To amend this, we propose to perform a high order energy conserving smoothing 
of the turbulence field obtained by the approximate de-convolution method 
\citep{Stolz:1999, Adams:2011, Hejlesen:2013, Hejlesen:2015d}. 
Using a Gaussian filter we obtain and $m$-th order filter from:
\begin{equation} 
\widehat{\zeta}(\boldsymbol{k}) = D_m \ \mbox{exp} \left( - \frac{\sigma^2 \, \boldsymbol{k}^2}{2} \right)
\qquad \mbox{with} \qquad
D_m = \displaystyle \sum_{n=0}^{m/2-1} \frac{(\sigma^2 \, \boldsymbol{k}^2/2)^n}{n!}
\end{equation}
where $D_m$ is the m-th order Taylor approximation of the inverse Gaussian, 
$\boldsymbol{k}^2 = \boldsymbol{k} \cdot \boldsymbol{k}$ and $\sigma$ is 
the smoothing radius which should 
not be smaller than the discretization length $h$. 
The smoothed turbulent vorticity field may now be obtained by:
\begin{equation}
 \widehat{\boldsymbol{\omega}}(\boldsymbol{k}) = \widehat{\zeta}(\boldsymbol{k}) \,  \widehat{\boldsymbol{\phi}}(\boldsymbol{k}) \cdot \widehat{\boldsymbol{H}}(\boldsymbol{k})
\label{eq-4:smooth generation}
\end{equation}
Once smoothed, we apply an additional correction to the generated vorticity field 
in order to enforce the constraint of a divergence-free vorticity field. 
From the kinematic relations of Eq.~\eqref{eq-4:kinematics} we may deduce that 
a vorticity field $\boldsymbol{\omega}_*$ 
where $\boldsymbol{\nabla} \cdot \boldsymbol{\omega}_* \neq 0$ should be 
corrected accordingly to: 
\begin{equation}
 \boldsymbol{\omega} = \boldsymbol{\omega}_* + \boldsymbol{\nabla} (\boldsymbol{\nabla} \cdot \boldsymbol{\psi}_*) 
 \qquad \mbox{and} \qquad
 \boldsymbol{\nabla} \cdot \boldsymbol{\omega}_* = - \boldsymbol{\nabla}^2 (\boldsymbol{\nabla} \cdot \boldsymbol{\psi}_*)
\end{equation}
The divergence of the uncorrected vector potential $\boldsymbol{\psi}_*$ is found by 
solving the latter equation which conveniently constitutes a Poisson equation 
and can thus be obtained using spectral differentiation. 
As seen in Eq.~\eqref{eq-4:kinematics} we also obtain 
the velocity field by solving the Poisson equation in a similar way.

Not only does the vorticity formulation of the turbulence generator provides a field 
with improved kinematic properties it also allows for an a priori determination 
of the Kolmogorov length which is the smallest length scale allowed by the 
viscous dissipation. By dimensional analysis we reformulate the conventional 
estimation of the Kolmogorov length as:
\begin{equation}
 \eta = \left( \frac{\nu^2}{ \overline{\varepsilon} } \right)^{1/4}
\end{equation}
where $\nu$ is the kinematic viscosity of the fluid,
$\varepsilon = \boldsymbol{\omega} \cdot \boldsymbol{\omega}$ is the enstrophy density 
and $\overline{\varepsilon}$ its mean value which 
can be determined directly from the generated field. The reformulated Kolmogorov 
length should be larger than the discretization length i.e. $\eta > h$ 
for a direct numerical simulation in order to avoid an excessive numerical dissipation. 
The Kolmogorov length should thus 
be used as the smoothing length $\sigma$ in order to ensure that the generated 
kinetic energy is contained in length scales which is allowed by the viscous 
dissipation of the fluid. 
If scales below the Kolmogorov length are introduced into a 
numerical simulation the viscous terms of the governing equations will dissipate 
the energy of these scales within a few time steps.

\section{Results}

\begin{figure}
\setlength{\unitlength}{0.0500bp}
\begin{picture}(8760,3276)
\put(0,0){
      \put(560,640){\makebox(0,0)[r]{\strut{}$10^{-6}$}}%
      \put(560,1438){\makebox(0,0)[r]{\strut{}$10^{-4}$}}%
      \put(560,2237){\makebox(0,0)[r]{\strut{}$10^{-2}$}}%
      \put(560,3035){\makebox(0,0)[r]{\strut{}$10^{0}$}}%
      \put(680,440){\makebox(0,0){\strut{}$10^{0}$}}%
      \put(2350,440){\makebox(0,0){\strut{}$10^{1}$}}%
      \put(4019,440){\makebox(0,0){\strut{}$10^{2}$}}%
      \put(0,1837){\rotatebox{-270}{\makebox(0,0){\strut{}Energy}}}%
      \put(2349,140){\makebox(0,0){\strut{}$k$}}%
    \put(-300,0){\includegraphics{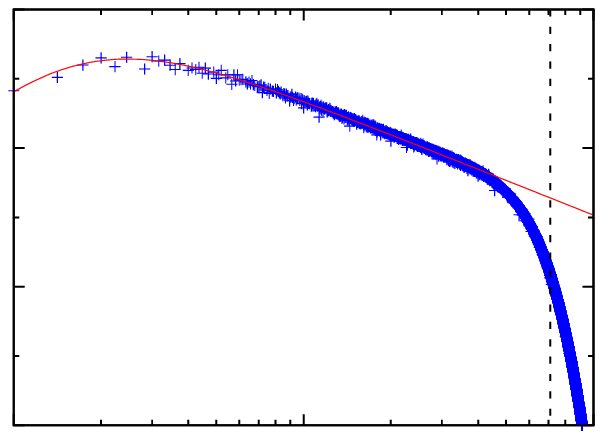}}%
}
\put(4680,0){    
      \put(560,640){\makebox(0,0)[r]{\strut{}$10^{-2}$}}%
      \put(560,1438){\makebox(0,0)[r]{\strut{}$10^{0}$}}%
      \put(560,2237){\makebox(0,0)[r]{\strut{}$10^{2}$}}%
      \put(560,3035){\makebox(0,0)[r]{\strut{}$10^{4}$}}%
      \put(680,440){\makebox(0,0){\strut{}$10^{0}$}}%
      \put(2350,440){\makebox(0,0){\strut{}$10^{1}$}}%
      \put(4019,440){\makebox(0,0){\strut{}$10^{2}$}}%
      \put(680,440){\makebox(0,0){\strut{}$10^{0}$}}%
      \put(2350,440){\makebox(0,0){\strut{}$10^{1}$}}%
      \put(4019,440){\makebox(0,0){\strut{}$10^{2}$}}%
      \put(0,1837){\rotatebox{-270}{\makebox(0,0){\strut{}Enstrophy}}}%
      \put(2349,140){\makebox(0,0){\strut{}$k$}}%
    \put(-300,0){\includegraphics{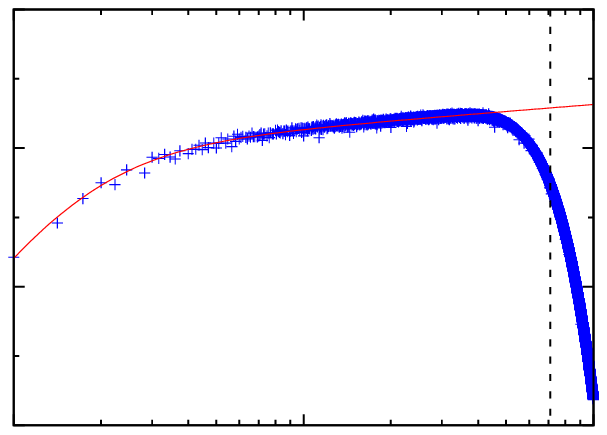}}%
}  
\put(0,3200){\makebox(0,0)[l]{(a)}}
\put(4680,3200){\makebox(0,0)[l]{(b)}}
\end{picture} 
\caption{
The energy spectrum (a) and the enstrophy spectrum (b) of a synthesized 
turbulence field generated on a $256 \times 256 \times 256$ grid 
using an integral length of $L = 0.1$ 
and a covariance tensor of $\boldsymbol{\Sigma} = \boldsymbol{I}$. 
The re-sampled spectra of stochastically generated turbulence 
field:~({\color{blue} \textbf{--------}}) is here compared to the 
target spectra used to generate the 
field:~({\color{red} \textbf{--------}}).
The wave-number of the smoothing 
radius $\sigma$:~({\color{black} \textbf{--~--~--~--}}) corresponding to 
minimum Kolmogorov length of $\eta/L = 0.0141$ or equivalently a 
kinematic viscosity of $\nu = 2.6 \times 10^{-3}$. 
A 10th order smoothing filter is used.
}
 \label{fig-4:energy spectra}
\end{figure}

The kinetic energy spectrum and the enstrophy 
spectrum of a generated isotropic turbulence field are shown in 
Fig.~\ref{fig-4:energy spectra}. 
The turbulent vorticity field is generated through Eq.~\eqref{eq-4:isotropic} 
using the von K{\'a}rm{\'a}n spectrum \citep{vonKarman:1948} for the 
turbulent kinetic energy $E(k)$. 
The von K{\'a}rm{\'a}n spectrum is here determined by defining 
an integral length $L$, and a scaling 
parameter $K = \rho \, \mbox{trace}(\boldsymbol{\Sigma})/2$ which is 
the total turbulent kinetic energy:
\begin{equation}
 E(k) = C \frac{L^4 k^4}{(1 + L^2 k^2)^{17/6}}
 \qquad \mbox{with} \qquad
 C = K \left( \int_0^\infty \frac{L^4 k^4}{(1 + L^2 k^2)^{17/6}} \ dk \right)^{-1}
\end{equation}
Hence we have the two input parameters $L$ and $\boldsymbol{\Sigma}$ 
representing the length scale of the largest coherent flow structures and 
the covariance tensor of the generated velocity field, respectively.
For the presented example the turbulent field of a cubed domain 
of $256 \times 256 \times 256$ cells is generated 
using $L = 0.1$ and $\boldsymbol{\Sigma} = \boldsymbol{I}$. 
The re-sampled spectra of the turbulence energy and enstrophy are shown in 
Fig.~\ref{fig-4:energy spectra}. The generated turbulence spectrum is found 
to converge to the design spectrum when averaging multiple realizations of the 
turbulence field. Furthermore, the calculated divergence in the velocity 
and the vorticity field was observed to be at the order of machine precision. 

\section{Conclusion}

A vorticity formulated stochastic turbulence generator was presented which 
improved the kinetic properties of the generated turbulent field compared 
to present methods. Additional measures, such as explicit high order smoothing 
of the flow field, was introduced to ensure that the generated field can be 
introduced into numerical simulations without an excessive loss of turbulence 
energy caused by numerical dissipation. The turbulence generator allows for 
a direct access to both the velocity and vorticity field without implementing 
supplementary schemes.


\section*{References}
\bibliographystyle{elsarticle-num}

\end{document}